\documentclass[prd,amsmath,amssymb,superscriptaddress,preprintnumbers,twocolumn,10pt]{revtex4-1}

\pdfoutput=1

\usepackage{graphicx}
\usepackage{dcolumn}
\usepackage{bm}
\usepackage{amssymb}
\usepackage{latexsym}
\usepackage{booktabs}
\usepackage{amsmath}
\usepackage{multirow}
\usepackage{url}

\usepackage{float}
\usepackage[colorlinks=true, linkcolor=red, citecolor=blue]{hyperref}

\usepackage[normalem]{ulem}
\usepackage{color}
\usepackage{array}
\usepackage{enumerate}

\newcommand{\aap}{Astron. Astrophys.}
\newcommand{\apjl}{Astrophys. J. Lett.}

\begin{document}

\title{
Analytical modeling of the one-dimensional power spectrum of 21-cm forest based on a halo model method
}

\author{Yue Shao}
\affiliation{Liaoning Key Laboratory of Cosmology and Astrophysics, College of Sciences, Northeastern University, Shenyang 110819, China}

\author{Tian-Yang Sun}
\affiliation{Liaoning Key Laboratory of Cosmology and Astrophysics, College of Sciences, Northeastern University, Shenyang 110819, China}

\author{Meng-Lin Zhao}
\affiliation{Liaoning Key Laboratory of Cosmology and Astrophysics, College of Sciences, Northeastern University, Shenyang 110819, China}
\affiliation{Theoretical Physics Division, Institute of High Energy Physics, Chinese Academy of Sciences, Beijing 100049, China}

\author{Xin Zhang}\thanks{Corresponding author}\email{zhangxin@mail.neu.edu.cn}
\affiliation{Liaoning Key Laboratory of Cosmology and Astrophysics, College of Sciences, Northeastern University, Shenyang 110819, China}
\affiliation{National Frontiers Science Center for Industrial Intelligence and Systems Optimization, Northeastern University, Shenyang 110819, China}
\affiliation{MOE Key Laboratory of Data Analytics and Optimization for Smart Industry, Northeastern University, Shenyang 110819, China}

\begin{abstract}
The 21-cm forest, composed of spectral absorption features from high-redshift background radio sources, provides a unique probe for studying small-scale structures during the epoch of reionization. It is particularly sensitive to detecting small-scale structures and early heating processes. Despite the rich information contained in the 21-cm forest signal, the complexity of directly modeling the signal has led to a lack of effective analytical models. However, the one-dimensional (1D) power spectrum of the 21-cm forest contains valuable information about the matter power spectrum, making analytical modeling feasible. This work employs an analytical modeling approach based on the halo model, which links the distribution of matter to dark matter halos, allowing for effective predictions of cosmic structure formation and its impact on the 21-cm signal. By considering various parameter scenarios within the halo model framework, particularly different dark matter particle masses and varying levels of cosmic heating, we can capture the complexities of small-scale structures and make the 1D power spectrum modeling applicable across a wide range of parameters. This method not only enhances our understanding of the 21-cm forest signal but also provides theoretical support for future observational data. Observing the 21-cm forest with large radio telescopes, such as the Square Kilometre Array, is anticipated to enable simultaneous exploration of dark matter properties and the heating history of the early Universe.
\end{abstract}

\maketitle

\newpage
\section{Introduction}

Originating from the hyperfine structure transition of neutral hydrogen atoms, the 21-cm signal constitutes a pivotal tool for investigating the first billion years of the Universe including the stages of cosmic dark ages, cosmic dawn, and the epoch of reionization (EoR). 
During the phase when reionization is still incomplete, a significant amount of neutral hydrogen atoms accumulate within and around small-scale structures such as minihalos and dwarf galaxies \cite{Furlanetto:2002ng,Xu:2010us,Meiksin:2011gx}. If the light from high-redshift radio-bright sources, such as radio-loud quasars \cite{Banados:2021imw,2021A&A...647L..11I,Haiman:2004ny,2022A&A...668A..27G} and the radio afterglows of gamma-ray bursts \cite{Ioka:2004zk,Salvaterra:2009ey,Cucchiara:2011pj,Ciardi:2015lia,Kinugawa:2019cgd}, happens to traverse these structures along the line of sight, the 21-cm absorption of neutral hydrogen will produce narrow and dense absorption lines in their spectra. This phenomenon is known as the ``21-cm forest'' \cite{Carilli:2002ky,Furlanetto:2002ng,Furlanetto:2006dt,Xu:2009dr,Xu:2010br,Xu:2010us,Ciardi:2012ik}.
The absorption line characteristics of the 21-cm forest can elucidate the small-scale distribution of neutral hydrogen throughout the EoR and yield vital insights into the heating processes and the formation of small-scale structures in the early Universe.


Dark matter (DM) is instrumental in the formation of cosmic structures. In contrast to the cold dark matter (CDM) model, non-CDM models, including warm dark matter (WDM) and ultralight dark matter, exhibit distinct behaviors on small scales \cite{Bode:2000gq,Marsh:2016vgj,Munoz:2019hjh,Jones:2021mrs,Ferreira:2020fam,Shen:2023lsf,Shimabukuro:2019gzu,Kawasaki:2020tbo}. Specifically, in the WDM model, the free-streaming of dark matter particles inhibits the formation of small-scale structures, resulting in a scarcity of dark matter halos \cite{Avila-Reese:2000nqd,Smith:2011ev,Schneider:2013ria}. Since the 21-cm forest signal is predominantly sourced by these low-mass dark matter halos, different dark matter models directly impact the abundance of 21-cm forest absorption lines \cite{Sitwell:2013fpa,Shimabukuro:2014ava}. Furthermore, early x-ray sources in the Universe heat the intergalactic medium (IGM), subsequently reducing the intensity and number of 21-cm signals \cite{Xu:2009dr,Mack:2011if,Shimabukuro:2014ava}. By measuring the one-dimensional (1D) power spectrum along the line of sight, it is feasible to disentangle these two effects, as their impacts on various scales are discernible. Consequently, the 21-cm forest serves as a dual probe for exploring the properties of dark matter and the heating levels in the early Universe \cite{Shao:2023agv}.

Despite the theoretical promise of the 21-cm forest, there remains a scarcity of analytical models for this probe. Direct modeling of the 21-cm forest is intricate due to the interplay of multiple physical processes, including dark matter halo formation, gas thermal evolution, and the effects of the radiation background \cite{Furlanetto:2002ng,Ciardi:2004ru,Xu:2010us,Soltinsky:2022jvf}. As upcoming observations become increasingly feasible, the lack of analytical models capable of linking observational data to key physical parameters such as the temperature of the IGM or the mass of WDM particles would make it necessary to rely on computationally expensive small-scale simulations. Therefore, it is crucial to develop simple yet effective analytical frameworks that can bridge observational quantities with underlying physical parameters. Conversely, the 1D power spectrum of the 21-cm forest encapsulates extensive information about the matter power spectrum, and the analytical modeling of matter power spectra is relatively mature \cite{Ma:2000ik,Smith:2002dz,Mead:2015yca,Philcox:2020rpe,Acuto:2021yjm}. Therefore, the analytical modeling of the 1D power spectrum presents a viable approach.

The halo model establishes a link between the distribution of matter and dark matter halos, providing effective predictions for cosmic structure formation \cite{Cooray:2002dia}. It has been extensively utilized in modeling the matter power spectrum, galaxy power spectrum \cite{Seljak:2000gq,Murray:2020dcd}, and 21-cm power spectrum \cite{Feng:2017ttq,Schneider:2020xmf,Schneider:2023ciq,Schaeffer:2023rsy,Hitz:2024cwl}, among other applications. For example, prior studies have employed the halo model to simulate the 21-cm power spectrum during the cosmic dawn, where overlapping radiation flux profiles delineate the impacts of Lyman-$\alpha$ (Ly$\alpha$) coupling and temperature fluctuations \cite{Schneider:2020xmf,Schneider:2023ciq,Schaeffer:2023rsy}. Additionally, comparisons between the halo model and the seminumerical code {\tt 21cmFAST} \cite{Mesinger:2010ne} have demonstrated good agreement on large scales, with discrepancies primarily emerging at small scales. {This work focuses on the application of the halo model to small-scale structures,}
as the 21-cm forest signal at small scales harbors rich information pivotal for studying dark matter properties and cosmic heating processes. Leveraging the high frequency resolution of radio telescopes such as the Square Kilometre Array (SKA), we can capture the physical processes within small-scale structures with greater precision.

Radio-loud quasars are the most suitable background sources for detecting the 21-cm forest. Numerous studies have already predicted the number of radio-loud quasars observable at high redshifts \cite{Haiman:2004ny,Wilman:2008ew,2021A&A...656A.137G}. Recently, Niu et al. \cite{Niu:2024eyf} updated the distribution of radio-loud quasars. Utilizing this updated methodology, we simulate the high-redshift radio-loud quasars observable by the SKA. We employ the halo model methodology, coupled with simulated high-redshift radio-loud quasars as background sources, to compute the 1D power spectrum of the 21-cm forest along the line of sight. We specifically consider the ramifications of different DM models (encompassing CDM and WDM) and varying heating levels (primarily x-ray heating). By juxtaposing our results with simulated data, we validate the reliability of our analytical model. This model facilitates an understanding of the perturbations in the 21-cm signal at small scales, thereby aiding in the exploration of dark matter properties and heating processes in the early Universe.

\section{Halo model formalism}\label{sec_hm}

\subsection{21-cm forest signal}

The 21-cm photons emitted by high-redshift background sources are absorbed by neutral hydrogen in and around minihalos (or dwarf galaxies), generating the 21-cm forest signal. The differential brightness temperature observed for the 21-cm signal is commonly expressed as
\begin{align}\label{dtb}
\delta T_{\mathrm{b}}(\hat{\boldsymbol{n}}, z) &\ \approx \frac{T_{\mathrm{S}}(\hat{\boldsymbol{n}}, z)-T_\gamma(\hat{\boldsymbol{n}}, z)}{1+z} \tau(\hat{\boldsymbol{n}}, z),
\end{align}
where $\hat{\boldsymbol{n}}$ represents the specific direction of the background source, $T_{\mathrm{S}}$ denotes the spin temperature of the HI gas and $\tau$ represents the 21-cm optical depth. 
$T_{\gamma}$ is the brightness temperature of the background radiation, including the background point source temperature $T_{\mathrm{point}}$ and the cosmic microwave background (CMB) temperature $T_{\mathrm{CMB}}$. 
The 21-cm optical depth, which quantifies the absorption strength, can be expressed in terms of the average gas properties as follows \cite{Furlanetto:2006dt,Madau:1996cs}:
\begin{equation}
\begin{aligned}\label{tau}
\tau(\hat{\boldsymbol{n}}, z) & \approx 0.0085 [1 + \delta(\hat{\boldsymbol{n}}, z)] (1 + z)^{3 / 2} \left[ \frac{x_{\mathrm{HI}} (\hat{\boldsymbol{n}}, z)}{T_{\mathrm{S}} (\hat{\boldsymbol{n}}, z)} \right] \\
& \times \left[ \frac{H(z) / (1 + z)}{\mathrm{d} v_{\|} / \mathrm{d} r_{\|}} \right] \left( \frac{\Omega_{\rm b} h^2}{0.022} \right) \left( \frac{0.14}{\Omega_{\rm m} h^2} \right).
\end{aligned}
\end{equation}
Here, $\delta(\hat{\boldsymbol{n}}, z)$ is the gas overdensity, $x_{\mathrm{HI}}(\hat{\boldsymbol{n}}, z)$ is the fraction of neutral hydrogen, $H(z)$ is the Hubble parameter, and $\mathrm{d} v_{\|} / \mathrm{d} r_{\|}$ is the velocity gradient along the line of sight. For the cosmological parameters, we set the values as follows: $h = 0.6736$, $\Omega_{\rm m} = 0.3153$, and $\Omega_{\rm b} h^2 = 0.02236$ \cite{Planck:2018vyg}. 

The observed brightness temperature of a background source, at a frequency $\nu$, is related to the flux density $S_{\mathrm{point}}(\nu)$ by 
\begin{align}
T_{\mathrm{point}}(\hat{\boldsymbol{n}}, \nu, z=0) = \frac{c^2}{2 k_{\rm B} \nu^2} \frac{S_{\mathrm{point}}(\nu)}{\Omega}.
\end{align}
Here, $c$ represents the speed of light, $k_{\rm B}$ is the Boltzmann constant, and $\Omega$ denotes the solid angle subtended by the telescope. The flux density $S_{\mathrm{point}}(\nu)$ of the background point source is modeled as a power-law spectrum with a spectral index $\eta$. This model is scaled to 150 MHz, represented as $S_{\rm point }(\nu) = S_{150} \left( \frac{\nu}{\nu_{150}} \right)^{\eta}$ \cite{Thyagarajan:2020nch}, where $S_{150}$ represents the flux density at $150 {\rm~MHz}$, with $\eta = -1.05$ \cite{Carilli:2002ky}.

Equations (\ref{dtb}) and (\ref{tau}) reveal that the 21-cm forest signal primarily depends on the gas density, spin temperature, neutral hydrogen fraction, and velocity gradient along the line of sight.
We assume $T_{\gamma} \gg T_{\rm S}$ and $T_{\rm S}$ is assumed to be fully coupled to the gas kinetic temperature $T_{\rm K}$ through the early Ly$\alpha$ background. $T_{\rm K}$ is determined by the heating history of the IGM, or the virial temperature of halos, depending on the gas location. For 21-cm forest observations, we only consider neutral patches, thus we can assume that $x_{\rm HI} = 1$. We can further simplify $\delta T_{\rm b}$ by
\begin{equation}
\begin{aligned}
\delta T_{\mathrm{b}}(\hat{\boldsymbol{n}}, z)  \approx  T_0(\hat{\boldsymbol{n}}, z) \frac{1+\delta(z)}{T_{\rm K}(z)},
\end{aligned}
\end{equation}
where $T_0(\hat{\boldsymbol{n}}, z) = - 0.0085 (1+z)^{1/2} T_{\gamma}(\hat{\boldsymbol{n}}, z)$ is only related to the brightness temperature of the background radiation. Although $T_{\rm K}$ and $\delta$ actually describe the temperature and density fields in the direction of the background source, we will show in the subsequent analytical modeling that these physical quantities essentially originate from the isotropic halo mass function and the density and temperature profiles. Therefore, they are statistically independent of direction. Based on this, we omit the direction $\hat{\boldsymbol{n}}$ in our notation and simply write them as $T_{\rm K}(z)$ and $\delta(z)$.

We utilize the 1D power spectrum along the line of sight to extract the scale dependence of the 21-cm forest signal, thereby enhancing the signal-to-noise ratio (SNR) and making the 21-cm forest an effective means \cite{Thyagarajan:2020nch,Shao:2023agv}. The 1D power spectrum of the 21-cm forest can be expressed as
\begin{align}\label{p21}
P(\hat{\boldsymbol{n}}, k_{\|}, z) = T_0^2(\hat{\boldsymbol{n}}, z) P_{21}(k_{\|}, z),
\end{align}
where $P_{21}(k_{\|}, z)$ is the 1D power spectrum of the term $[1+\delta(z)]/T_{\rm K}(z)$. It should be noted that $k_\parallel$ is, by definition, the Fourier variable of the comoving distance $r_z(z)$ along the line of sight. Therefore, it should theoretically vary with redshift $z$. However, in practical power spectrum analysis, due to the finite length of the observed or simulated region, the $k$ modes obtained from the discrete Fourier transform have statistical errors, especially in the high-$k$ region where large fluctuations are more likely to occur. 
To mitigate such uncertainties, it is common in power spectrum analysis to apply binning in $k$ space, where adjacent $k$ modes are grouped and averaged within the same bin. This procedure reduces statistical fluctuations in the estimated power spectrum, resulting in more reliable measurements. As a result, the binned $k_\parallel$ no longer corresponds to a specific redshift, but instead represents the average power over a fixed scale interval.
Therefore, in subsequent analyses, we treat it as a redshift-independent variable to facilitate the unified comparison of power spectra and parameter constraints across different redshifts. We can obtain the 1D power spectrum by projecting the averaged three-dimensional (3D) power spectrum onto a specific direction,
\begin{align}\label{pave}
P_{21}(k_{\|}, z) = P_{21,{\rm 1D}}(k, z) = \frac{1}{2 \pi} \int_k^{\infty} k' P_{21}(k', z) {\rm d}k'.
\end{align}
It should be noted that the 3D power spectrum $P_{21}(k, z)$ here is determined solely by the spatial distribution of neutral hydrogen, which depends on the density profile, temperature profile, and halo mass function of dark matter halos. In our model, these quantities are assumed to be isotropic, and thus $ P_{21}(k, z)$ is also isotropic. In addition, redshift space distortions mainly originate from the peculiar velocities of the gas, which only cause a slight frequency drift in the absorption signal. We consider that the impact of this on the overall amplitude of the 1D power spectrum is relatively weak. Therefore, it is neglected in this work. Meanwhile, since we focus on scales smaller than 2 Mpc, the evolution of physical quantities along the line of sight with redshift can also be considered negligible on this scale. Based on these premises, we can project it along any direction to obtain the 1D power spectrum $P_{\rm 21,1D}(k, z)$, and the result along the line of sight is $P_{21}(k_\parallel, z)$. Finally, through Eq.~(\ref{p21}), we obtain the 1D power spectrum of the 21-cm forest $P(\hat{\boldsymbol{n}}, k_{\|}, z)$ along the line of sight. The information related to the direction of the background radio source including background source flux density $S_{150}$ is encoded in $T_0^2(\hat{\boldsymbol{n}},z)$.

\subsection{Halo model}

Following the computation of the 1D power spectrum, we now introduce the halo model formalism used to calculate the averaged 3D power spectrum of the 21-cm forest signal. The halo model provides a framework for analytically describing the clustering of matter on both small and large scales by dividing the power spectrum into contributions from individual dark matter halos (the one-halo term) and correlations between separate halos (the two-halo term).

The averaged 3D power spectrum $P_{21}(k, z)$ consists of two components,
\begin{align}
P_{21}(k, z) = P_{21}^{1h}(k, z) + P_{21}^{2h}(k, z),
\end{align}
where $P_{21}^{1h}(k, z)$ represents the contribution from a single halo and $P_{21}^{2h}(k, z)$ captures the contribution from two separate halos. The specific expressions for each term are given as follows:
\begin{equation}
\begin{aligned}
P_{21}^{1h}(k, z) &= \frac{1}{\langle \rho_{21} \rangle^2} \int {\rm d} M \frac{{\rm d} n}{{\rm d} M} |W_{21}|^2, \\
P_{21}^{2h}(k, z) &= \frac{1}{\langle \rho_{21} \rangle^2} 
\left[ \int {\rm d} M \frac{{\rm d} n}{{\rm d} M} |W_{21}| b \right]^2 \times P_{\rm lin}.
\end{aligned}
\end{equation}
Here, ${\rm d} n(z, M) / {\rm d} M$ is the halo mass function \cite{Press:1973iz,Barkana:2000fd}, $b(z, M)$ is the halo bias \cite{Mo:1995cs}, and $P_{\rm lin} (k, z)$ is the linear matter power spectrum.
The window function $W_{21}(k, z, M)$ represents the Fourier transform of the profile $\rho_{21}(r, z, M)$, given by
\begin{align}
W_{21}(k, z, M) = 4 \pi \int {\rm d} r r^2 \rho_{21}(r, z, M) \frac{\sin(k r)}{k r}.
\end{align}
The mean density profile $\langle \rho_{21}(z) \rangle$ is obtained by integrating over the halo mass function and the profile radii,
\begin{align}
\langle \rho_{21}(z) \rangle = 4 \pi \int {\rm d} M \frac{{\rm d} n(z, M)}{{\rm d} M} \int {\rm d} r r^2 \rho_{21}(r, z, M),
\end{align}
where $\rho_{21}(r, z, M)$ is the profile of $[1+\delta(z)]/T_{\rm K}(z)$, including both the density profile and the temperature profile. In the following sections, we will expand on these components, detailing the specific forms of the density and temperature profiles.

\subsubsection{Halo mass function}

The halo mass function is particularly crucial for modeling how these halos contribute to the 21-cm forest signal, especially for low-mass halos found in neutral regions. In the CDM model, the number density of halos within a certain mass range at redshift $z$ is described by the Press-Schechter formalism \cite{Press:1973iz,Barkana:2000fd}.  The halo mass function is expressed as
\begin{align}
\frac{{\rm d} n(z, M)}{{\rm d} M} = \sqrt{\frac{2}{\pi}} \frac{\bar{\rho}_{\rm m0}}{M} \left|\frac{{\rm d} \sigma(M)}{{\rm d} M}\right| \frac{\delta_{\rm c}(z)}{\sigma^3(M)} \exp \left[ -\frac{\delta^2_{\rm c}(z)}{2\sigma^2(M)} \right],
\end{align}
where $\bar{\rho}_{\rm m0}$ is the current average matter density in the Universe, $\sigma(M)$ is the standard deviation of the mass distribution, and $\delta_{\rm c}(z)$ is the critical overdensity for collapse at redshift $z$.

\begin{figure}
\centering
\includegraphics[angle=0, width=8cm]{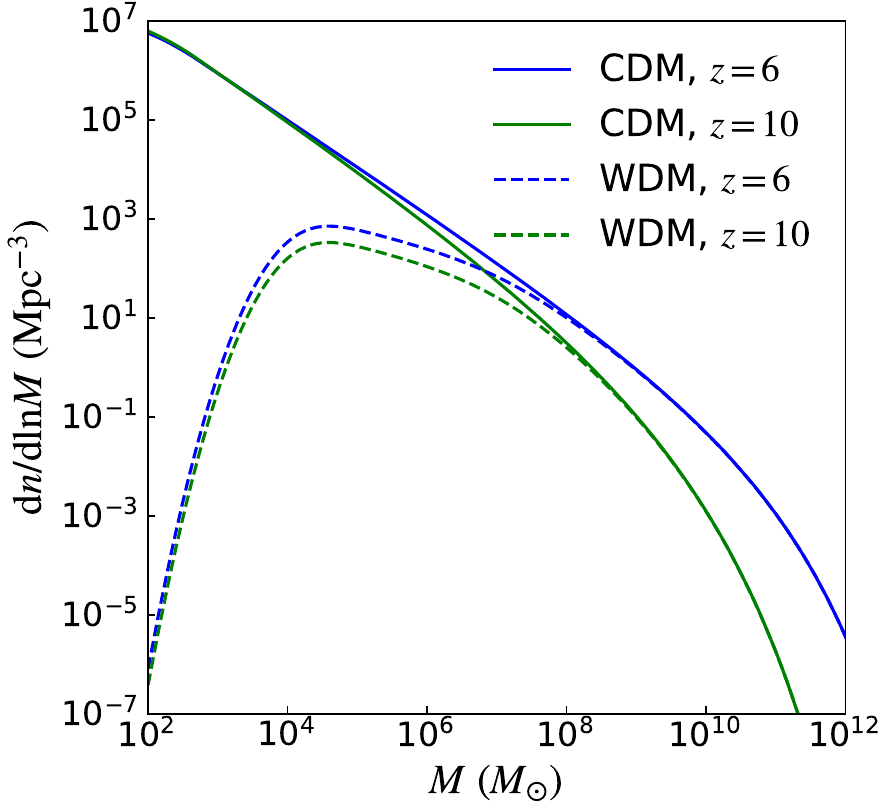}
\caption{\label{mass_func}
  Halo mass function in CDM model (solid lines) and WDM model (dashed lines) with $m_{\rm W} = 6 {\rm ~keV}$ at different redshifts. The blue and green lines correspond to $z$ = 6 and 10, respectively. 
}
\end{figure}

In contrast, the WDM model shows suppressed structure formation below the free-streaming scale $\lambda_{\rm fs}$, reflecting the thermal velocities of WDM particles. This effect leads to a decrease in the number of low-mass halos. The WDM halo mass function is approximated by \cite{Smith:2011ev}
\begin{align}
\frac{{\rm d} n(z, M)}{{\rm d} M} = \frac{1}{2} \left\{ 1 + \operatorname{erf} \left[ \frac{\log_{10}(M / M_{\rm fs})}{\sigma_{\log M}} \right] \right\} \left[ \frac{{\rm d} n(z, M)}{{\rm d} M} \right]_{\rm PS},
\end{align}
where $\sigma_{\log M} = 0.5$ is the width of the transition and $ M_{\rm fs}$ is the characteristic mass scale corresponding to the free-streaming scale $\lambda_{\rm fs}$. The term $\left[ {\rm d} n(z, M)/{\rm d} M \right]_{\rm PS}$ is the Press-Schechter mass function from the CDM model.
The free-streaming scale for WDM particles is given by \cite{Smith:2011ev}
\begin{align}
\lambda_{\rm fs} \approx 0.11 \left( \frac{\Omega_{\rm W} h^2}{0.15} \right)^{1 / 3} \left( \frac{m_{\rm W}}{\mathrm{keV}} \right)^{-4 / 3} \, \text{Mpc},
\end{align}
where $\Omega_{\rm W}$ is the WDM density relative to the critical density and $m_{\rm W}$ is the mass of the WDM particle. Current constraints on the mass of WDM particles are typically on the order of a few keV \cite{Palanque-Delabrouille:2019iyz,Garzilli:2019qki,Enzi:2020ieg,Villasenor:2022aiy,Dayal:2023nwi}. Smaller WDM particle masses lead to a greater suppression scale, substantially decreasing the number of small halos and affecting the structure of the 21-cm forest signal.
The Press-Schechter mass function for WDM involves modifying the CDM power spectrum to account for suppression on small scales. The matter power spectrum for WDM is given by \cite{Bode:2000gq}
\begin{align}
P_{\rm WDM}(k) = P_{\rm CDM}(k) \left[ \left( 1 + (\alpha k)^{2 \beta} \right)^{-5 / \beta} \right]^2,
\end{align}
where $\beta$ = 1.12 and $\alpha$ is a scale factor that depends on the WDM particle mass \cite{Viel:2005qj},
\begin{align}
\alpha = 0.049 \left( \frac{m_{\rm W}}{\mathrm{keV}} \right)^{-1.11} \left( \frac{\Omega_{\rm W}}{0.25} \right)^{0.11} \left( \frac{h}{0.7} \right)^{1.22} h^{-1}\text{Mpc}.
\end{align}
Figure~\ref{mass_func} illustrates the halo mass functions for CDM and WDM models at different redshifts. Since the 21-cm forest signal is mainly contributed by low-mass halos, we focus on halos with a minimum mass of $M_{\rm min} = 10^5 M_\odot$, corresponding to the Jeans mass scale for the redshifts of interest. The upper mass limit, $M_4$, is associated with a virial temperature of $T_{\rm vir} = 10^4 {\rm ~K}$. Within this mass range, the WDM model predicts fewer halos than the CDM model.

\subsubsection{Density profile}

The distribution of gas density within and around dark matter halos plays a crucial role in shaping the 21-cm absorption signal, as it directly influences the distribution of neutral hydrogen along the line of sight. For regions inside the virial radius $r_{\rm vir}$, we assume that the dark matter adheres to the Navarro-Frenk-White (NFW) \cite{Navarro:1996gj} density profile, while the gas maintains hydrostatic equilibrium with the dark matter distribution. Based on these assumptions, the gas density profile can be analytically derived as \cite{Makino:1997dv}
\begin{align}
\ln \rho_{\rm g}(r) = \ln \rho_{\rm gc} - \frac{\mu m_{\rm p}}{2 k_{\rm B} T_{\rm vir}} \left[ v_{\rm e}^{2}(0) - v_{\rm e}^{2}(r) \right],
\end{align}
where $\rho_{\rm g}(r)$ is the gas density at distance $r$ from the halo center, $\rho_{\rm gc}$ is the central gas density, $\mu$ is the mean molecular weight of the gas, $m_{\rm p}$ is the proton mass and $v_{\rm e}(r)$ is the gas escape velocity at radius $r$.
The expression for the escape velocity $v_{\rm e}(r)$ is as follows,
\begin{align}
v_{\rm e}^{2}(r) = 2 \int_{r}^{\infty} \frac{G M(r')}{r'^{2}} \, {\rm d}r' = 2 V_{\rm c}^{2} \frac{F(yx) + \frac{yx}{1 + yx}}{x F(y)},
\end{align}
where $V_{\rm c}^2 \equiv G M / r_{\rm vir}$ is the circular velocity at the virial radius,
$x = r / r_{\rm vir}$ is the normalized radius, $F(y) = \ln(1 + y) - y / (1 + y)$, and $G$ is the gravitational constant. 
$y$ denotes the halo concentration parameter, which correlates with the halo mass and is commonly known as the core-halo mass relation. We utilize the fitting relationship presented in Ref.~\cite{Gao:2005hn}. Although this relation may vary across different dark matter models, for simplicity, we apply the same core-halo mass relation across all DM models.

Because of the gravitational influence, the gas density in the surroundings of halos is enhanced. Outside the virial radius, we assume the gas density profile aligns with that of the dark matter and can be determined using the infall model \cite{Barkana:2002bm}. The gas density profiles in and around halos of different masses are plotted in the left panel of Fig.~\ref{profile}.

\subsubsection{Temperature profile}

The temperature of the gas, especially in the neutral IGM, is vital in influencing the 21-cm absorption signal as it impacts both the spin temperature of hydrogen and the absorption strength. The temperature of the gas within the neutral IGM is mainly regulated by cosmic expansion, Compton scattering with the CMB, and x-ray heating. While ultraviolet radiation primarily heats ionized regions, x-rays can deeply penetrate the neutral IGM, serving as the main heating source for the gas responsible for the 21-cm signal. The overall temperature evolution of the IGM, denoted as $T_{\rm K}$, can be described by \cite{Furlanetto:2006tf}
\begin{align}
\frac{{\rm d} T_{\rm K}}{{\rm d} t} = -2 H(z) T_{\rm K} + \frac{2}{3} \frac{\epsilon_{\rm comp}}{k_{\rm B} n} + \frac{2}{3} \frac{\epsilon_{\rm X,h}}{k_{\rm B} n},
\end{align}
where $n$ is the total particle number density, $\epsilon_{\rm comp}$ is the Compton heating/cooling rate per unit volume and $\epsilon_{\rm X,h}$ is the x-ray heating rate.
The term $\epsilon_{\rm X,h}$ represents the portion of the x-ray emissivity that contributes to heating, which depends on the ionized fraction $x_i$. A fitting formula gives $\epsilon_{\rm X,h} = [1 - 0.8751(1 - x_i^{0.4052})] \epsilon_{\rm X}$ \cite{Valdes:2008cr}.
The total x-ray emissivity $\epsilon_{\rm X}$ is proportional to the star formation rate, which is linked to the collapse rate of matter into halos. This can be written as \cite{Furlanetto:2006tf}
\begin{align}
\frac{2}{3} \frac{\epsilon_{\rm X}}{k_{\rm B} n H(z)} = 5 \times 10^{4} \, {\rm K} \, f_{\rm X} \left( \frac{f_\star}{0.1} \frac{{\rm d} f_{\rm coll} / {\rm d} z}{0.01} \frac{1 + z}{10} \right),
\end{align}
where $f_\star$ is the star formation efficiency, $f_{\rm coll}$ is the fraction of matter collapsed into atomic-cooling halos, and $f_{\rm X}$ is a normalization factor representing the uncertain x-ray productivity in the early Universe.

Within the virial radius $r_{\rm vir}$, the gas kinetic temperature $T_{\rm K}$ is equated to the virial temperature $T_{\rm vir}$ of the halo. The virial temperature is essential for determining the ability of the gas to cool and form stars. In the regions surrounding halos, gas is adiabatically heated based on the local density, with temperature profiles increasing toward the halo center.
In the regions outside the virial radius, the gas temperature is determined by a balance between adiabatic cooling due to cosmic expansion and heating from x-rays. In the presence of significant x-ray heating, the gas temperature is set by the maximum of the adiabatic temperature and the x-ray heated IGM temperature. This ensures that gas far from halos remains warmer if x-rays dominate the heating process.

\begin{figure}
\centering
\includegraphics[angle=0, width=8.5cm]{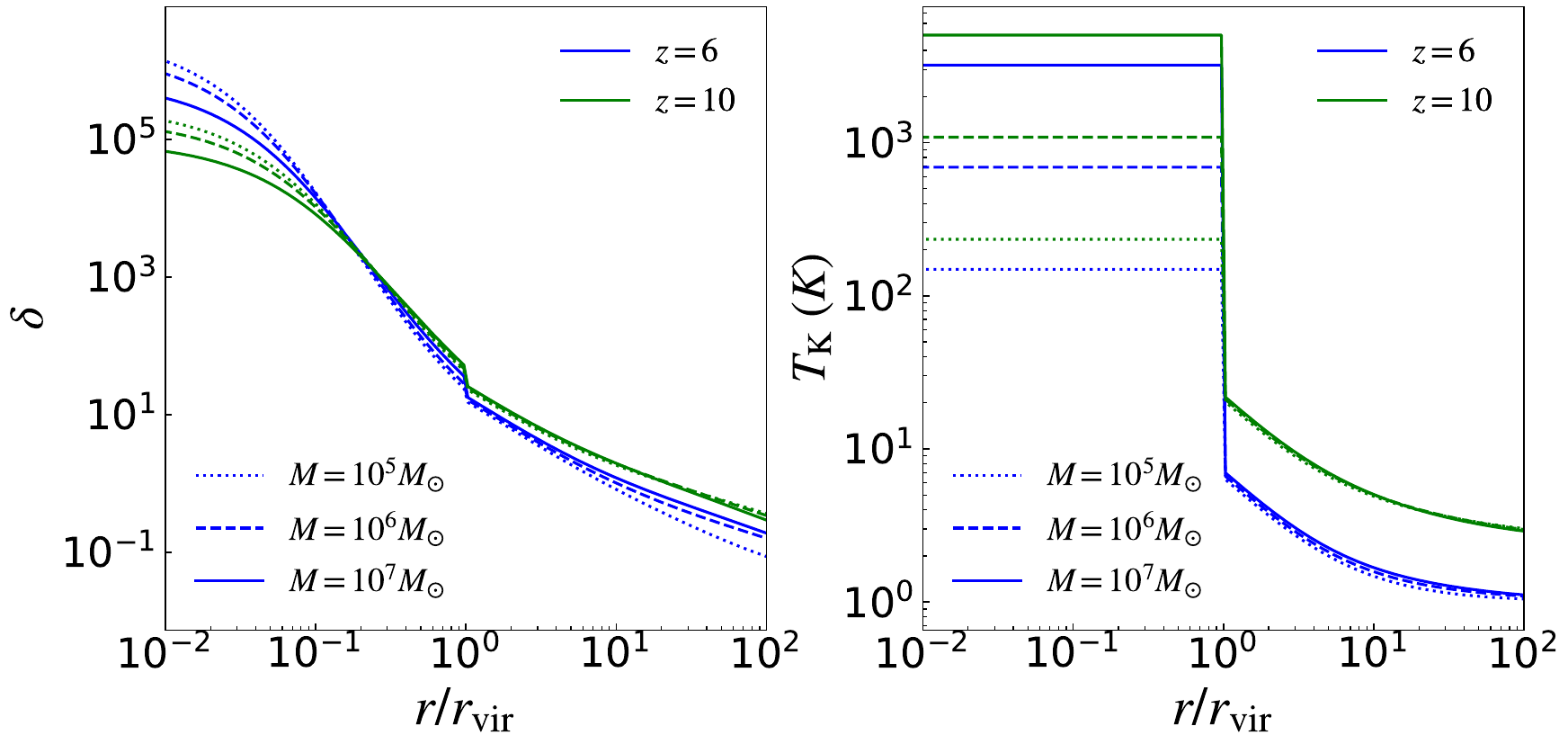}
\caption{\label{profile}
  The neutral hydrogen overdensity profiles (left panel) and temperature profiles (right panel) inside and outside the virial radius of a halo.
  In both panels, the blue and green lines correspond to $z$ = 6 and 10, respectively. The dotted, dashed and solid lines correspond to halo mass of $10^5 M_{\odot}$, $10^6 M_{\odot}$ and $10^7 M_{\odot}$, respectively.
}
\end{figure}

The heating due to x-rays is dependent on the parameter $f_{\rm X}$, which characterizes the efficiency of x-ray production from early sources. For higher values of $f_{\rm X}$, the gas temperature in neutral regions is significantly raised, reducing the contrast between dense regions and the IGM in terms of their 21-cm absorption signatures. We show the temperature profiles for an unheated IGM with different masses in the right panel of Fig.~\ref{profile}.
According to the recent Hydrogen Epoch of Reionization Array findings~\cite{HERA:2022wmy}, the temperature of IGM at redshift $z \sim $ 8 is constrained within the range $15.6 {\rm ~K} < T_{\rm K} < 656.7 {\rm ~K}$, corresponding to $0.02 < f_{\rm X} < 0.6$. Thus for subsequent analysis, we adopt $f_{\rm X} = 0.1$ as the fiducial model.

\section{Forecasts}

Overall, the 1D power spectrum modeled using the halo model method is primarily influenced by the halo mass function, density profile, and temperature profile. The properties of DM and the temperature of IGM are the main factors affecting these components, corresponding to the WDM particle mass $m_{\rm W}$ and the heating level $f_{\rm X}$. In this section, we briefly introduce the simulation of background sources. We use Bayesian analysis methods, based on the 1D power spectrum of the 21-cm forest from these background sources, to evaluate the ability of halo model to constrain $m_{\rm W}$ and $f_{\rm X}$.

\subsection{Radio-loud quasar population}
{With the advancement of optical and near-infrared detection technology, the frontier of quasar redshift has reached 7.64 \cite{2021ApJ...907L...1W}, while about $10\%$ of quasars exhibit strong radio emission \cite{Stern:2000kc,SDSS:2002iri,Banados:2015fda,2021A&A...656A.137G,2021ApJ...908..124L}, making these radio-loud quasars ideal candidates for 21-cm forest observations. Using a physics-driven model \cite{Haiman:2004ny,Niu:2024eyf}, we can predict the future observations of radio-loud quasars. In combination with future infrared space telescopes, such as Euclid \cite{Euclid:2021icp} and the Nancy Grace Roman Space Telescope \cite{Spergel:2015sza}, we will be able to  discover more high-redshift active galactic nuclei (AGNs). Additionally, by utilizing future powerful radio telescopes like the SKA, the New Extension in Nançay Upgrading Low-Frequency Array \cite{Munshi:2023buw}, and the upgraded Giant Metrewave Radio Telescope \cite{2017CSci..113..707G}, we can filter out high-redshift radio-loud quasars from these AGNs.
}

We utilize the latest model of radio-loud quasars distribution presented in Ref.~\cite{Niu:2024eyf} to simulate background point sources.
The abundance of radio-loud quasars can be calculated as follows:
\begin{equation}
\begin{aligned}
\frac{\mathrm{d} n(z, F_{\rm th})}{\mathrm{d} z \mathrm{d} \Omega} 
& =\frac{\mathrm{d} V}{\mathrm{d} z \mathrm{d} \Omega} D_{\mathrm{q}}(z)  \\
& \times \int \mathrm{d}M \frac{\mathrm{d} n(z,M)}{\mathrm{d} M} 
\int_{R_0(M,F_{\rm th})} \mathrm{d} R N(R) ,
\end{aligned}
\end{equation}
where $\mathrm{d} V / \mathrm{d} z \mathrm{d} \Omega$ is the cosmological volume element, $D_{\rm q}$ is the duty cycle and $R_0$ is the threshold of radio loudness, determined from the radio flux density threshold $F_{\rm th}$. $N(R)$ is the radio-loudness distribution of the radio-loud quasars,
\begin{align}
N(R) = \frac{1}{\sqrt{2\pi}\sigma_0} \exp \left[\frac{-(R-\bar{R})^2}{2\sigma_0^2} \right],
\end{align}
where $R$ represents radio loudness. $\bar{R} = 2.67$ and $\sigma_0 = 0.55$ represent the best fitting parameters obtained from Ref.~\cite{Niu:2024eyf}.

To calculate the survey area of SKA-LOW, we apply the following formula:
\begin{align}
S_{\rm area} = \eta \int \sin\theta {\rm d}\theta \int {\rm d} \phi,
\end{align} 
where $\eta$ represents the fraction of the initial survey area that remains usable for observation after excluding the Milky Way, for which we adopt a value of 0.7. Considering a declination range from $34^\circ$ to $-86^\circ$, the total survey area is approximately 10,000 deg$^2$.

The noise flux density observed by SKA-LOW can be estimated as \cite{2017isra.book.....T}
\begin{equation}
\delta S^{\rm N} \approx \frac{2 k_{\mathrm{B}} T_{\mathrm{sys}}}{A_{\mathrm{eff}} \sqrt{2 \delta \nu \delta t}},
\end{equation}
where $\delta \nu$ is the channel width, $\delta t$ is the integration time, $A_{\rm eff}$ is the effective collecting area of the telescope and $T_{\rm sys}$ is the system temperature. In our analysis, we use $\delta \nu = 1 {\rm ~kHz}$ and $\delta t = 100 {\rm ~h}$. We adopt $A_{\rm eff} / T_{\rm sys}=600 \mathrm{~m}^2 \mathrm{~K}^{-1}$ for the SKA-LOW \cite{Acedo:2020lve,Braun:2019gdo}. We consider SNR greater than 5 as the criterion for detecting radio-loud quasars. Assuming an optical depth $\tau = 0.1$, to meet this criterion, we need to satisfy the condition $F_{\rm th} \times (1-e^{-\tau}) > 5 \times \delta S^{\rm N}$.

\begin{figure}
\centering
\includegraphics[angle=0, width=8cm]{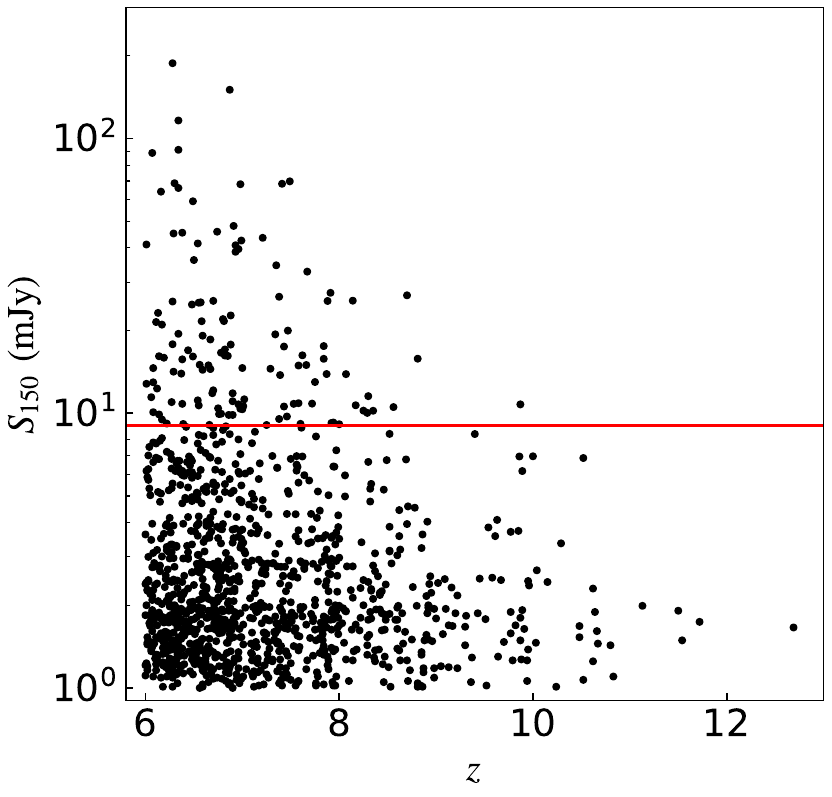}
\caption{\label{qso}
  Simulated radio-loud quasar distribution with SKA-LOW. The red line represents the flux threshold that SKA-LOW can observe.
}
\end{figure}

We present the $z-S_{150}$ distribution of radio-loud quasars in Fig.~\ref{qso}. Assuming an integration time of $\delta t = 100 {\rm ~h}$ and a channel width of $\delta \nu = 1 {\rm ~kHz}$, under this configuration, quasars with a flux $S_{150} \gtrsim 10 {\rm ~mJy}$ at low redshifts between 6 and 8 can all be observed, and there are also a few quasars with a flux $S_{150} \gtrsim  100 {\rm ~mJy}$. As the redshift increases, only a small number of quasars with a flux above $10 {\rm ~mJy}$ can be detected. Particularly, it becomes very difficult to find suitable background sources for observation at redshifts above 10. For these fainter quasars, observations can be improved by increasing the integration time or decreasing the frequency resolution. However, lower frequency resolution reduces the number of measurements, which in $k$ space means fewer $k$ modes, leading to increased sample variance and thus greater uncertainty in the measurements.
From these, we select 10 radio-loud quasars with significant flux densities at different redshifts for further analysis.

\subsection{Parameter estimation}

The observational uncertainties in the 21-cm forest include the thermal noise and the sample variance. The thermal noise uncertainty for the 1D power spectrum is given by \cite{Thyagarajan:2020nch}
\begin{align}
P^{\rm N}=\left(\frac{\lambda_z^2 T_{\mathrm{sys}}}{ A_{\rm {eff }} \Omega}\right)^2\left(\frac{\Delta r_z}{2 \Delta \nu_z \delta t_{0.5}}\right),
\end{align}
where $\Delta \nu_z$ is the total observing bandwidth corresponding to $\Delta r_z$, $\delta t_{0.5}$ is half of the integration time. Here, $\Delta r_z$ denotes the comoving length of the observed segment along the line of sight. If $N_s$ neutral segments can be observed, the thermal noise will be reduced by a factor of $\sqrt{N_s}$. Here, $N_s$ represents the total number of usable neutral segments along the sightlines of all observed quasars, which depends on the number of background sources and the number of neutral regions that can be segmented in the spectrum of each background source. In actual observations of the 21-cm forest signal, we can divide the total observation time into two parts and perform cross-correlation on the signals observed in these two segments. This approach helps to reduce the thermal noise. Similarly, in our work, we employed this method to estimate the thermal noise. 
We utilize $P (k_{\|}, z)$ to estimate the sample variance, and the total measurement error of the 1D power spectrum $\sigma_{P}$ is
\begin{align}
\sigma_{P} = \frac{1}{\sqrt{N_{\rm m}}} \left[ P^{\rm N} + P (k_{\|}, z) \right],
\end{align}
where $N_{\rm m}$ is the number of $k$ modes, which includes the number of source spectra segment and the number of $k$ modes in each $k$ bin. 

Because of current observations only placing a lower limit on $m_{\rm W}$, we employ Bayesian analysis to deduce its reciprocal during the inference process. We employed {\tt Bilby} \cite{Ashton:2018jfp} for Bayesian analysis to estimate the parameters $f_{\rm X}$ and $m_{\rm W}^{-1}$. For the observed 1D power spectrum $\tilde{P}$, it can be expressed as
\begin{align}
\mathcal{P}(f_{\rm X} , m_{\rm W}^{-1} \mid \tilde{P})=\frac{\mathcal{P}(\tilde{P} \mid f_{\rm X} , m_{\rm W}^{-1}) \mathcal{P}( f_{\rm X} , m_{\rm W}^{-1})}{\mathcal{P}(\tilde{P})},
\end{align}
where $\mathcal{P}(f_{\rm X} , m_{\rm W}^{-1} \mid \tilde{P})$ is the posterior distribution, $\mathcal{P}(\tilde{P} \mid f_{\rm X} , m_{\rm W}^{-1})$ is the likelihood function, $\mathcal{P}( f_{\rm X} , m_{\rm W}^{-1})$ is the prior distribution, and $\mathcal{P}(\tilde{P})$ is the normalization constant. The likelihood function for the 1D power spectrum $\tilde{P} _{i,k}$ , based on the $i$th observation and the $k$th bin, is defined as follows:
\begin{equation}
\begin{aligned}
\mathcal{P}(\tilde{P} \mid f_{\rm X} , m_{\rm W}^{-1})
 = \prod_{i,k} 
\frac{1}{\sqrt{2 \pi \sigma_{P_{i,k}}^{2}}}
\exp \left[-\frac{\left(\tilde{P} _{i,k}-P_{i,k}\right)^{2}}{2 \sigma_{P_{i,k}}^{2}}\right],
\end{aligned}
\end{equation}
where $\sigma_{P_{i,k}}$ is the measurement error. The prior distributions for the parameters $f_{\rm X}$ and $m_{\rm W}^{-1}$ are uniform distributions over the range $[0.01, 1]$ and $[1/9, 1/3]~\rm{keV}^{-1}$. We used the Markov chain Monte Carlo \cite{Ashton:2021anp} sampling algorithm to draw samples from the posterior distribution, estimating the parameters $f_{\rm X}$ and $m_{\rm W}^{-1}$ and their precision.

\section{Results and discussion}\label{res}

\begin{figure}
\centering
\includegraphics[angle=0, width=8.5cm]{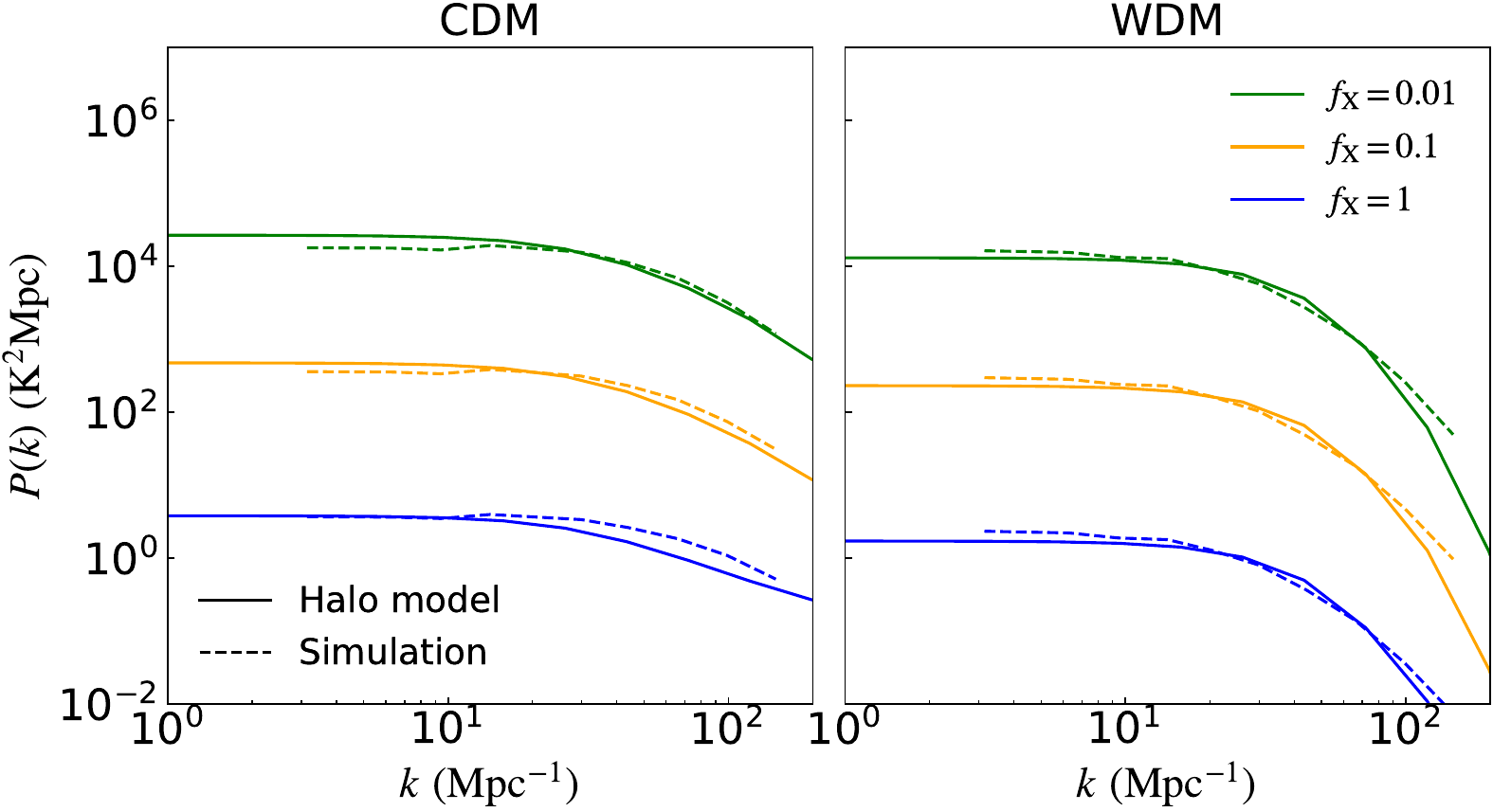}
\caption{\label{halomodel}
  Comparison of the 21-cm forest halo model presented in this paper (solid curves) with the simulation results (dashed curves) at $z = 9$, assuming a flux density of $S_{150} = 10 {\rm ~mJy}$. The green, yellow and blue curves correspond to $f_{\rm X}$ = 0.01, 0.1 and 1, respectively. The left and right panels represent CDM model and WDM model with $m_{\rm W} = 6 {\rm ~keV}$, respectively.
}
\end{figure}

While the DM model and heating effects suppress the 21-cm forest signal, their impacts on different scales vary significantly. These differences can be revealed through the 1D power spectrum. To elucidate these scale-dependent effects, we developed an analytical model for the 1D power spectrum of the 21-cm forest using the halo model method and conducted detailed calculations of the 1D power spectra under various DM models and heating levels.
We consider a distance of 2 comoving Mpc along the line of sight, corresponding to a bandwidth of $\Delta \nu_z=0.11~\mathrm{MHz}$ and $k_{\rm min} = 3.14~{\rm Mpc}^{-1}$ at $z = 10$. 
Here, we only consider modeling on small scales below $2 {\rm ~Mpc}$ for comparison with the simulation result. 
Additionally, modeling larger scales that correspond to a broader bandwidth requires consideration of the ionized fraction, which adds complexity to the model. For some radio-loud quasars at relatively low redshifts, especially those near redshift 6 during the late stages of reionization, the ionization fraction of the Universe has already increased significantly. As a result, neutral regions are separated by large ionized regions. Against this backdrop, it is still feasible to extract short neutral segments from quasar spectra for the analysis of the 21-cm forest. However, extracting longer lines to probe larger-scale structures becomes very challenging due to the presence of ionized regions. The ionization field not only affects the length and distribution characteristics of the neutral segments but also impacts the 1D power spectrum of the 21-cm forest. Therefore, to extend our analytical model to larger scales, it is necessary in future studies to incorporate the effects of ionization into the modeling framework.
On the other hand, a larger bandwidth also increases the number of accessible $k$ modes, which can help reduce sample variance in statistical measurements.

For a direct comparison between our analytical model and the simulated 1D power spectra from previous studies, we conducted small-scale simulations on grids of $2 {\rm ~Mpc}$ in length \cite{Shao:2023agv}. Each $(2 {\rm ~Mpc})^3$ grid is populated with halos based on the conditional halo mass function for the CDM or WDM \cite{Cooray:2002dia,Zentner:2006vw}. The initial density is the mean density of the Universe. Specifically, within halos, the gas temperature is set to the virial temperature, while in the surrounding IGM, it is determined by adiabatic cooling or x-ray heating. The heating rate is regulated by an efficiency parameter $f_{\rm X}$, which controls the strength of x-ray emission from early x-ray sources.
The density in each voxel is determined by the NFW profile or the infall model profile. The gas temperature within each voxel is determined by its location relative to halos and the thermal history of the Universe. 
The large-scale reionization history only determines the probability of encountering a neutral patch of the IGM of a certain length along the line of sight, and we focus only on neutral regions on small scales. We assume that gas within and around low-mass halos is in collisional ionization equilibrium. Since the ionization fraction in such regions remains very low, we neglect ionization effects and consider only fully neutral patches.
After generating the density and temperature fields in the simulation, we further assume different background source flux densities to compute the 21-cm forest brightness temperature and the corresponding 1D power spectrum.
Each grid is divided into $500^3$ voxels, corresponding to a voxel size of approximately 4 kpc, and we average over $500^2$ lines along the line of sight direction to obtain the final 1D power spectrum for comparison with the analytical model, thereby mitigating the impact of sample variance. 
It is worth noting that the multiscale simulations presented in Ref.~\cite{Shao:2023agv} combine large-scale reionization and density fields from {\tt 21cmFAST} with small-scale modeling of halo properties. In contrast, our analytical model focuses solely on small-scale structures. Therefore, for consistency, we compare our results only with the small-scale parts (i.e., scales below 2 Mpc) of the simulations in Ref.~\cite{Shao:2023agv}.

In Fig.~\ref{halomodel}, we present a comparison between the 21-cm forest halo model developed in this work and the simulation results obtained using methods from Ref.~\cite{Shao:2023agv}. We display the results for both the CDM model (left panel) and the WDM model (right panel) at different heating levels, with fixed redshift $z = 9$ and $S_{150} = 10 {\rm ~mJy}$. 
Our analytical model shows good agreement with simulation results across various DM models and different levels of x-ray heating.
It is noteworthy that, at such small scales, the one-halo term dominates the 1D power spectrum. Therefore, we only present the results for the one-halo term.

\begin{figure}
\centering
\includegraphics[angle=0, width=8.5cm]{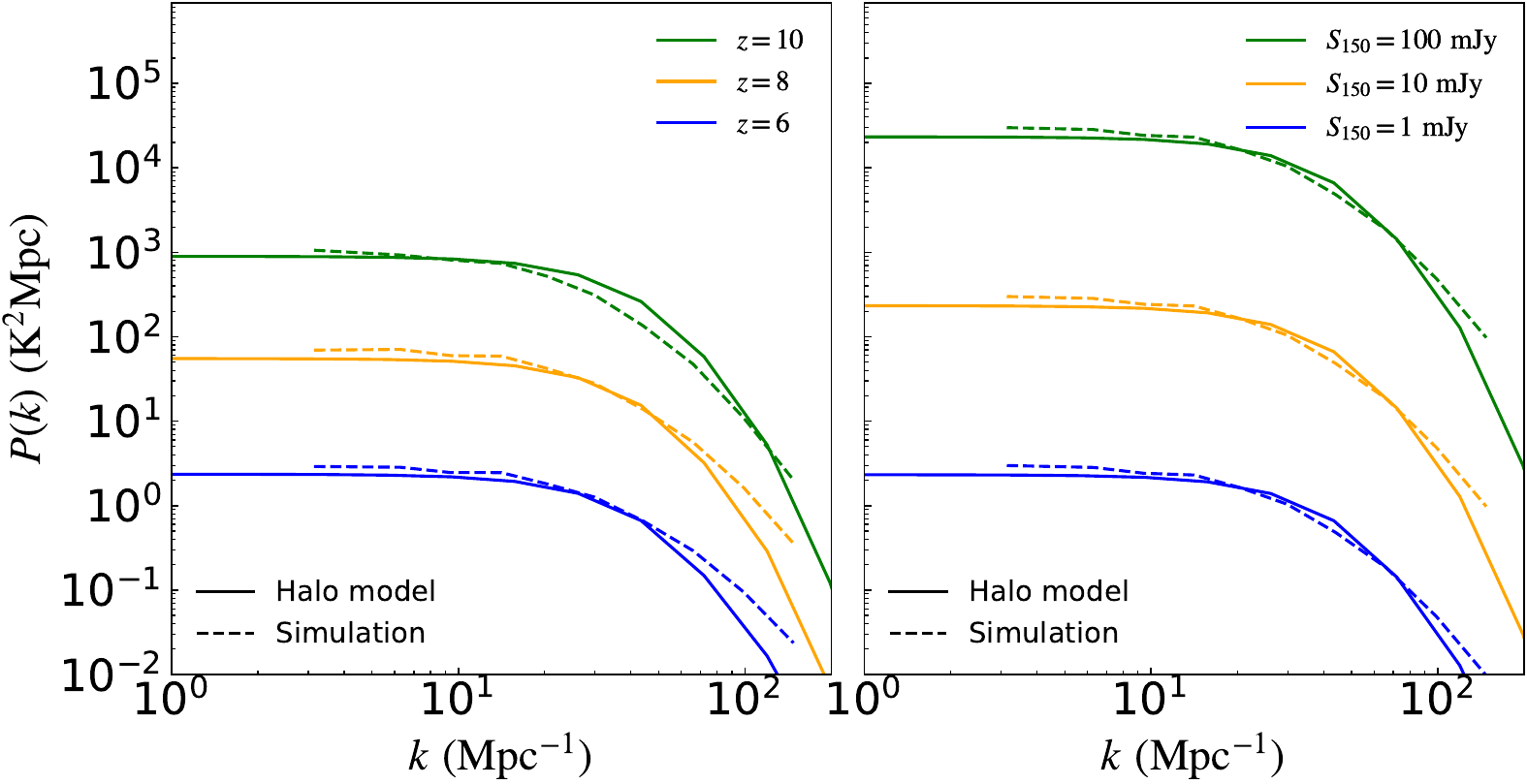}
\caption{\label{halomodel2}
  Comparison between the 21-cm forest halo model in this paper (solid curves) and the simulation results (dashed curves).
  The blue, yellow, and green curves in the left panel correspond to $z$ = 6, 8, and 10, respectively, with $S_{150} = 10 {\rm ~mJy}$. The blue, yellow, and green curves in the right panel correspond to $S_{150} = 1$, $10$, and $100 {\rm ~mJy}$, respectively, at $z = 9$.
}
\end{figure}

We fix $m_{\rm W} = 6 {\rm ~keV}$ and $f_{\rm X} = 0.1$ as the fiducial model, and calculate the performance of the 21-cm forest halo model at different redshifts and with different background source flux densities in Fig.~\ref{halomodel2}. The left panel shows the results for redshifts $z$ = 6, 8, and 10, assuming a uniform flux density $S_{150} = 10 {\rm ~mJy}$ for consistency. Note that in practice, at low redshifts, many quasars can have a flux density exceeding $10 {\rm ~mJy}$, whereas at high redshifts, such quasars are relatively few. The right panel fixes the redshift $z = 9$ and calculates the results for different background source flux densities $S_{150} = 1$, $10$, and $100 {\rm ~mJy}$, respectively. This indicates that our model is applicable across a wide range of parameter space.

\begin{figure}
\centering
\includegraphics[angle=0, width=8.5cm]{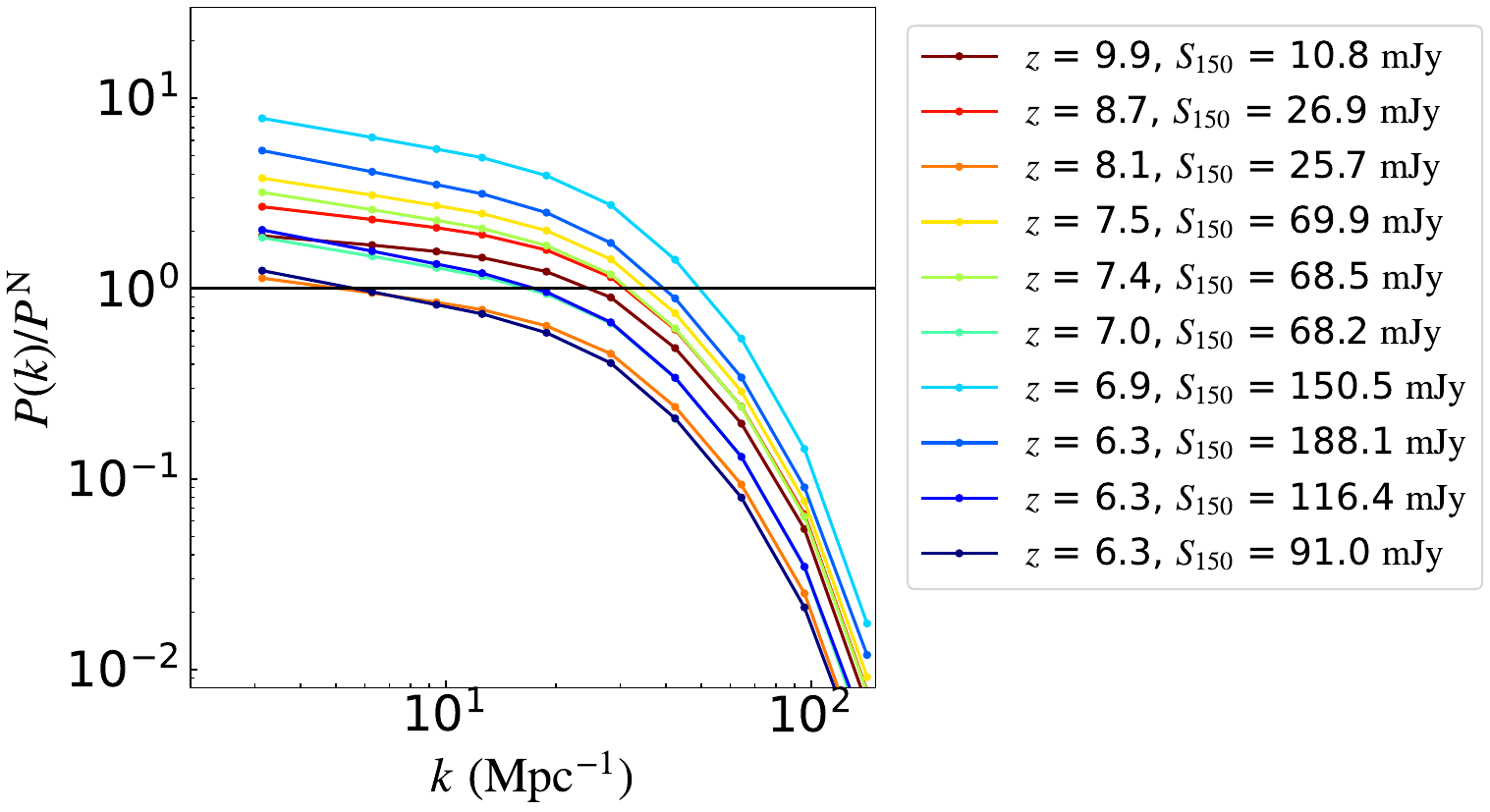}
\caption{\label{ps1d}
  The SNR, $P(k)/P^{\rm N}$, of the 21-cm forest calculated from 10 radio-loud quasars. Each quasar has 10 segments along the line of sight and an observation time of $\delta t = 100~ {\rm h}$.
}
\end{figure}

To accurately describe the heating history of the Universe and constrain the heating level $f_{\rm X}$, we aimed to identify the most promising radio-loud quasars for 21-cm forest observations. We first computed the total signal-to-noise ratio, $\sum P(k)/P^{\rm N}$, of the 1D power spectrum of the 21-cm forest for all radio-loud quasars detectable by SKA. Based on this, we selected the 10 sources with the highest total SNRs for our analysis. This selection reflects both an idealized and realistic scenario: under limited observational resources, prioritizing brighter quasars with higher SNR allows for more effective constraints on key physical parameters, such as the mass of warm dark matter particles and the level of cosmic heating. Such an approach is not only theoretically motivated but also aligns with realistic time-allocation considerations in actual observations. 
For each of these 10 quasars, we assume the presence of 10 neutral segments along the line of sight, each with a comoving length of 2 Mpc. The resulting SNRs of the 1D power spectra are shown in Fig.~\ref{ps1d}. Considering that thermal noise varies with redshift, we present the SNR in the form $P(k)/P^{\rm N}$.
At higher redshifts, the temperature of the IGM is lower, which means that quasars can achieve sufficient SNR even with moderate flux densities. In contrast, at lower redshifts, quasars need to be brighter to attain the same level of SNR. Quasars at redshifts between 7 and 8 are particularly well suited for observations of the 1D power spectrum of the 21-cm forest.

At the end of the EoR, the neutral hydrogen fraction of the Universe is already quite low, and most of the Universe at $z \sim 6$ has been ionized. {However, numerous observations suggest that the end of reionization occurred later than anticipated, possibly close to $z \sim 5.3$ \cite{Kulkarni:2018erh,Nasir:2019iop,Qin:2021gkn}. Therefore, it is also possible to detect a significant amount of 21-cm forest signals during the late stages of reionization \cite{Soltinsky:2021esh}.} Furthermore, since the lengths of the segments we are considering are very short, only about $2 {\rm ~Mpc}$, it is still possible to have up to ten segments of neutral regions.
Additionally, for radio-loud quasars at higher redshifts, there may be many more neutral segments. To simplify, we assume that each quasar has 10 segments, totaling 100 segments.
We take the redshift of the quasars as the redshift of the 21-cm forest signal and assume that the quasars do not produce a heating effect.

\begin{figure}
\centering
\includegraphics[angle=0, width=8.5cm]{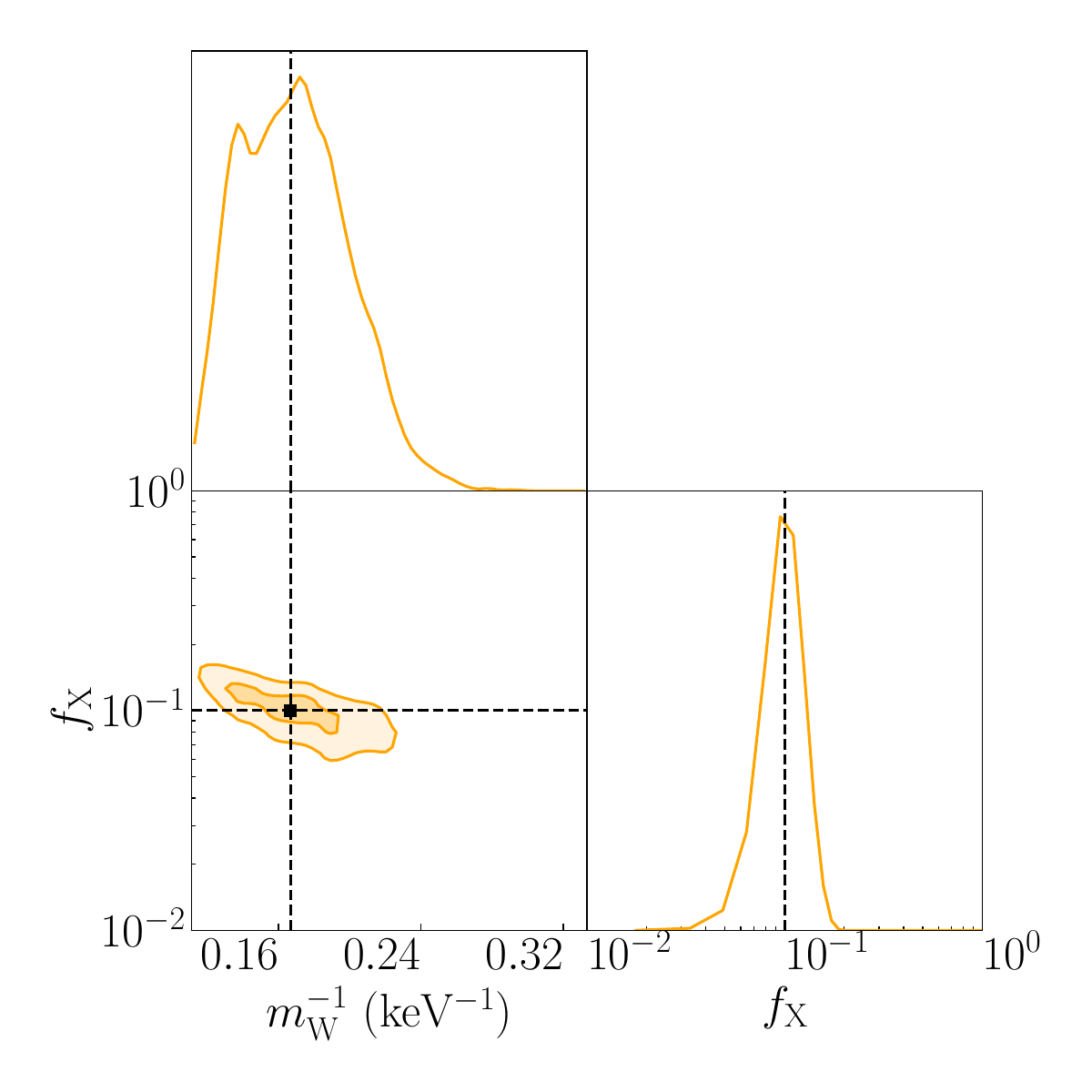}
\caption{\label{contour}
  Bayesian constraints at the $68.3\%$ and $95.4\%$ confidence levels for the parameters $f_{\rm X}$ and $m_{\rm W}$, derived from the 1D power spectrum of the 21-cm forest observed with SKA-LOW.
}
\end{figure}

We applied the Bayesian methods to constrain the parameters $m_{\rm W}$ and $f_{\rm X}$, which are presented in Fig.~\ref{contour}. Based on the prospective observations with SKA-LOW, we assumed 100 h of observation for each quasar, accumulating to a total of 1000 h of observational time.
We adopted a fiducial model with $m_{\rm W} = 6 {\rm ~keV}$ and $f_{\rm X} = 0.1$, and calculated the absolute errors for the parameters to be $\sigma_{m_{\rm W}} = 1.3 {\rm~keV}$  and $\sigma_{f_{\rm X}} = 0.02$. By combining 21-cm forest data from various redshifts, we can effectively constrain the mass of WDM particles and trace the evolution of the cosmic thermal history, thereby revealing the formation of the first galaxies. 

Here, we employ Bayesian methods as they offer distinct advantages over Fisher matrix analysis in handling complex probability distributions and the nonlinear nature of 1D power spectrum data. Furthermore, potential non-Gaussian features within the 21-cm data can be addressed by incorporating more complex likelihood functions or deep learning approaches \cite{Sun:2024ywb} to capture intricate patterns in the data.

Observing the 21-cm forest is contingent upon high-redshift, radio-bright background sources, with radio-loud quasars and radio afterglows of gamma-ray bursts emerging as the most viable options. As some of the brightest objects in the Universe, quasars' radio emissions can serve as the background for the 21-cm forest. In particular, radio-loud quasars provide sufficient intensity for observing the 21-cm signal. 
Using the quasar luminosity function, it is possible to predict the number of quasars, and in the SKA era, it is expected that a sufficient number of radio-loud quasars will be detected to serve as background sources for the 21-cm forest \cite{Banados:2021imw,2021A&A...647L..11I,Haiman:2004ny,2022A&A...668A..27G}. Some models have suggested that, although the abundance of radio-loud quasars at high redshifts is low, it is still adequate to support 21-cm forest studies \cite{2021A&A...656A.137G,Niu:2024eyf}.

It is commonly believed that galaxies can form only in dark matter halos with masses above $10^8 M_\odot$. However, some studies have suggested that Pop III galaxies might also form in low-mass halos with masses below $10^8 M_\odot$ via molecular cooling. The ultraviolet and x-ray radiation produced by these galaxies can heat and ionize the IGM, thereby affecting the statistical properties of the 21-cm signal \cite{Pochinda:2023uom,Lazare:2023jkg}. Nevertheless, the halo model adopted in this paper is primarily designed for the statistical modeling of the 21-cm forest absorption signal structure and does not yet account for the potential formation of galaxies in low-mass halos and their radiative feedback. The formation of molecular-cooling galaxies in low-mass halos and their radiative processes, especially their effects on the 21-cm forest signal and the 1D power spectrum in particular, are important topics that require further in-depth investigation in the future.

\section{Conclusion}
In this work, we propose an innovative analytical modeling approach, using the halo model method for the first time to theoretically modeling the 1D power spectrum of the 21-cm forest. This method not only addresses the lack of analytical models for the 21-cm forest but also significantly reduces the need for simulations. 
By comparing our model with the results from commonly used small-scale simulation methods for the 21-cm forest, we find that our model exhibits high consistency with the simulation results across a wide range of parameter spaces. This consistency not only validates the effectiveness of our model but also demonstrates its universality under different parameter conditions.

The 21-cm forest signal primarily comes from the neutral hydrogen within and around halos, so we focus on small-scale modeling. However, modeling at larger scales is also important. For high-redshift radio point sources, the neutral segments in their light paths are numerous and long, making $2 {\rm ~Mpc}$ scale modeling inadequate. For low-redshift radio point sources, modeling at larger scales requires considering the  ionized fraction of neutral hydrogen, which helps explore the evolution of the reionization history.

By incorporating effects such as ionization fluctuations and Ly$\alpha$ coupling, the halo model method can also be applied to larger scales. Alternatively, other possible analytical models can be developed \cite{Mirocha:2022pys,Munoz:2023kkg}. With the help of these analytical models, future studies related to the 21-cm forest can more accurately predict and interpret observational data, making the 21-cm forest a reliable probe for studying small-scale structures in the early Universe. This probe will not only help us better understand the nature of dark matter but also reveal the heating history of the early Universe.

\section*{Acknowledgments}
{We would like to thank Yidong Xu for the early-stage suggestion to use the halo model for analytical modeling, for the insightful discussions regarding the halo model method, and for providing the code about small-scale simulation of 21-cm forest.} Additionally, we also thank Yichao Li and Qi Niu for the fruitful discussions.
This work was supported by the National SKA Program of China (Grant No. 2022SKA0110200 and No. 2022SKA0110203), the National Natural Science Foundation of China (Grant No. 12533001, No. 12473001, No. 11975072,  No.11875102, and No. 11835009), and the National 111 Project (Grant No. B16009).

\section*{Data Availability}
The data are not publicly available. The data are available from the authors upon reasonable request.

\bibliography{main}

\end{document}